\documentclass[runningheads]{cl2emult}

\usepackage{makeidx}  
\usepackage{graphicx} 
\usepackage{subeqnar} 
\usepackage{multicol} 
\usepackage{cropmark} 
\usepackage{eso}      
\makeindex            

\begin{document}

\title*{The Space Missions  WAXS/WFXT and SWIFT: 
New Targets for the VLT}

\toctitle{The Space Missions  WAXS/WFXT and SWIFT: 
New Targets for the VLT}

\titlerunning{The Space Missions  WAXS/WFXT and SWIFT}
\author{Guido Chincarini}

\authorrunning{Guido Chincarini\inst{1}\inst{2}}

\institute{Osservatorio Astronomico di Brera, via E. Bianchi 46,
     I-23807 Merate (LC), Italy
\and Universit\`a degli Studi di Milano, via Celoria 6, I-20133
     Milano, Italy}

\maketitle             

\begin{abstract}
At OAB we were, during the year 1998, deeply involved in planning two
important space missions for which very large ground based telescopes,
VLT in particular, would play a very large and important role in the
optical followup. The study of the first mission, Wide Angle X-ray
Survey using a Wide Field X-ray Telescope, was coordinated by the
Observatory of Brera, involved mainly Italian industries and resulted
in a proposal to the Italian Space Agency under the Small Satellite
Program. The second mission, SWIFT, has been coordinated and directed
by the Goddard Space Flight Center and resulted in the submission of a
proposal to NASA under the MIDEX program. The science goal of this
mission is the detection and study of the Gamma-ray bursts.

\end{abstract}

\section{Background of the WAXS/WFXT Mission}
\vskip -0.3cm

It is very well known that the clue to the understanding the formation
and evolution of Clusters of Galaxies resides in the detection and
detailed study of clusters at redshifts larger than 0.5 and up to a
redshift z$\sim 2$ assuming they still exist and are recognizable. It is
also well known that the detection of distant clusters by optical
surveys is very limited because of the confusion of the galaxy
background at faint magnitudes and the fading in that of the member
galaxies. Indeed that had been clearly discussed some time ago by Cappi
et al. (1989) and more recently evidenced by the galaxy surveys carried
out by Postman et al. (1996) and by the ESO Survey (EIS), these
proceedings.
On the other hand while the X-ray cluster detection is a much cleaner
method due to the net separation at any redshift between an extended
emission and a point like source over a rather faint X-ray background, a
good resolution of the optics is a must. Possibly the X-ray telescope
should allow a resolution better than 15 arcsec Half Energy Width (HEW).
A good sensitivity is also needed since the surface brightness of even
bright clusters dims considerably with distance.
It has been shown, especially by Bahcall and Soneira (1983) but see
also the most recent results obtained with the REFLEX project, Guzzo
et al.  (1999), that clusters are unique tools in describing the
statistical characteristics of the Large Scale Structure. While
shortly the Power Spectrum for the local Universe will be quite nicely
measured by REFLEX, it remains fundamental to verify if the
theoretical prediction are confirmed at redshifts z$\sim 1$ and
larger.  Statistical information on much higher redshifts could be
obtained only by a deep AGN sample at z$ > 2.5$. To create a
statistical significant catalogue we need to survey a very large and
fairly deep area of the sky.  Following ROSAT observations and
hydrodynamics simulations, furthermore, it has been shown that groups
of clusters are interconnected by reasonably hot matter and galaxies
which may extend, at lower temperature however, in the Intergalactic
Medium forming an interconnected net spreading about the Universe in
the form of filaments (see Chincarini and Rood 1980, and references
therein, for an early discussion of this concept). Clusters, during
their lifetime, will be acreting matter, gas and galaxies, from these
filaments and from the surrounding environment and this will
characterize their properties at various redshifts. It becomes very
important, therefore, to have a rather large field of view in order to
observe simultaneously, and with comparable good resolution, the
clusters and their close environment.
To optimize this research sometime ago we started to design, and plan,
an X-ray space mission capable of optimizing the achievement of the
science goals. The requirements, derived directly by what has been
briefly mentioned above, were: 1) X-ray optics with good resolution
over a large field of view, 2) reasonable good sensitivity over the
energy range $0.2 - 2.5$ keV with some sensitivity up to about $7-8$
keV and 3) high accuracy in the image reconstruction (better than 2
arcsec) and absolute astrometry better than $2 - 3$ arcsec, this being
very important for AGN especially.  This idea, originally triggered by
Riccardo Giacconi, grew interest in the international astronomical
community and we started long ago to refine the details and study the
spacecraft and instrument requirements.  The Astronomical Observatory
of Brera got deeply involved not only in the definition of the Science
but, with the support of ASI and under the technical direction of
Citterio, started to develop and refine the design of a wide field
optics and its construction. More recently the effort of a rather
large international collaboration came up with a detailed study of
phase A submitted to ASI by the end of 1998. I refer for a detailed
discussion of the science and of the project to the proposal, and
references therein, we submitted (about 250 pages) and to the paper
presented by Chincarini (1999) on behalf of the collaboration.  In
this brief communication I will limit myself to describe the very
recent results obtained with the last optics we made and measured at
the Marshall Space Flight Center X-ray facility since this, by itself,
is a major result in the fabrication of high quality X-ray
optics. This is a break through equivalent to what occurred when we
discovered the design, and had the capability of making, the Ritchey
Chretien mirrors in place of the standard parabola and hyperbole.
\vskip -0.1cm

\section{The performance of the X-ray Optics}
\vskip -0.3cm

The results we are going to illustrate have been obtained with a 60 cm
wide shell made with a carrier of Silicon Carbide. The optical surface
is achieved by filling the gap between the super-polished mandrin
and the carrier with epoxy resin. The reflecting surface is made out
of gold.  The fastest way to illustrate the result is by Figure 1. The
two horizontal lines indicate the minimum acceptable resolution for
the mission ($\leq 20$ arc sec) and the goal ($\leq 15$ arc sec). The
two continuous lines at the bottom of the Figure refer to the optical
design of the mirror. The design of the optics is obtained by
optimizing the resolution over about 1 degree field of view and that
is the input for the construction of the super-polished mandrin whose
resolution, after construction, is given by the continuous line
running between 6 and 8 arc sec HEW. Obviously no replica can be
better than the master mandrin.  To illustrate the performance of a
classical Wolter I optics we dashed the resolution as a function of
the distance from the optical axis for the XMM telescope. As it is
well known while on axis and within 10 arc minutes the resolution is
good, it deteriorates drastically at larger distance from the optical
axis. That is over a field of 1 square degree the ratio (Area with
good resolution)/(Area with poor resolution) $\sim 0.16$. Above 20 arc
min off center the resolution is not even acceptable.  The WFXT optics,
the various replica have been obtained using always the same mandrin,
have been tested at the PANTER facility of MPE (Germany) and at the
Marshall facility (USA). The filled points refer to the first shell at
1.5 keV energy while the filled square refer to the tests obtained at
0.5 keV. This is already a great result! It shows, but we had already
some tests showing this, that we are very close to reaching the target
goal and that we are capable of making optics having good resolution
over a large field of view. We gain, compared to the Wolter I optics
above, $84\%$ of the field! Excellency, however, was reached testing a
new shell, Sic2.  In the making of this we obviously used of the
manufacturing experience we gained with the previous ones.  Here at
0.1 keV (we can measure at this low energy only at Marshall) we
measured an HEW of about 10 arc sec, stars in Figure 1 (due to a minor
miss-alignment the measurement at -30 arc min is actually more than
30' off axis). This test was very significant for an other aspect. The
quality of the images deteriorates somewhat going to higher energy due
to the development of wings in the profile which are due to
scattering.  By reducing the number of replica per mandrin and by
caring somewhat more in the process, that can be strongly reduced.

\begin{figure}
\centering
\includegraphics[width=.65\textwidth, angle=-90]{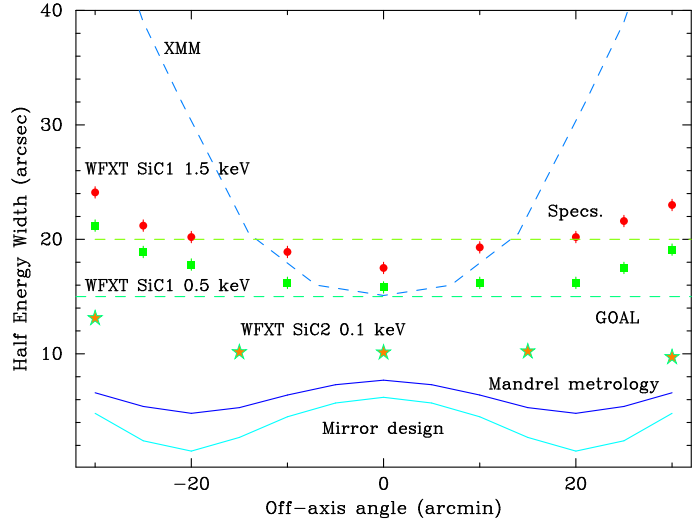}
\caption[]{Resolution of the X-ray optic vs. off-axis angle}
\label{eps1}
\end{figure}

With this result and with the phase A study on the focal plane (SAO
and Leicester), telescope and spacecraft we have shown that we are
ready for a fantastic mission because of the goals we can achieve in
cosmology and in other astronomical studies and because of
readiness. Whether it will fly or not will depend on the budget and on
the vision, or strong limitations and cultural bias, of those who will
evaluate it.
\vskip -0.1cm

\section{SWIFT - A panchromatic Gamma Ray Burst MIDEX Mission}
\vskip -0.3cm

In the limited time, and space, at my disposal I can not make justice
to such an excellent mission. I feel however obliged to mention it,
more information can be obtained from {\it
http://swift.gfsc.nasa.gov/}, because in the coming years I have no
doubt that the VLT will play a fundamental role in the follow up
studies of these very energetic events. A second motivation is that
during this meeting, Gamma ray bursts were not mentioned at all.  

The discovery by BeppoSAX of afterglow from Gamma-ray bursts has
revolutionized our understanding of these events and opened an
extremely interesting field of endeavor which will keep busy the
science community for years to come. We are dealing with extreme
energetic events, 10$^{51}$ - 10$^{54}$ ergs, which occur, in a very
short time. The understanding of the phenomenon requires detailed
observations of the burst itself and, whenever possible, of the faint
hosting galaxy.  It is of paramount importance to observe at various
wavelengths soon after the burst has been detected. SWIFT satisfies to
this need. The best way to illustrate the capability of the planned
mission is to mention briefly the mode of operation.  The spacecraft
host three instruments. A Burst Alert Telescope (BAT) sensitive in the
range 10 - 150 keV and capable of determining the GRB location to
about 4 arc minutes. It has a detecting area of 5200 cm$^2$ and a
Field of View of about 2 sr. An X-Ray telescope (XRT) sensitive in the
range 0.2-10keV and capable of locating the afterglow with an accuracy
of $\sim 2$ arc sec. The effective area at 1.5 keV is 110 cm$^2$. An
Ultra Violet Optical telescope (UVOT) which can observe in 6 colors
(from 170 nm to 650 nm). The aperture of the telescope is 30 cm with a
pixel scale of 0.5 arc sec.

The winning strategy is fast response and communication. Soon after
the burst (10 sec) the slew begins and the BAT location is transmitted
to Earth. After about 55 sec in addition to having the BAT observation
we plan to have the XRT location and image which, within 10 sec, is
send to Earth together with the BAT light curve. If visible by UVOT
this instrument will gather data in parallel with the other instrument
and send a $2\times 2$ arc minutes$^2$ finding card to the ground. It
will take about 1200 sec to get the XRT spectra and 7200 sec to have a
set of UVOT observations in the different filters.  Etc. The strategy
and the mission design satisfy the two main requirements mentioned
above: immediate accurate pointing of the spacecraft after the
detection of the burst and immediate alert to the ground to enable the
medium size and large telescopes to initiate the follow-up
observations. In this role the VLT, and other ground-based telescopes,
will play a fundamental role in measuring those quantities that allow
the understanding and modeling of the phenomenon.  Since these objects
are at cosmological distances, indeed it may be a way to know about
the Universe at very large redshifts, the observations of the host
galaxy, when detected, is a must. These observations must be carried
out with 4 m and 8 m class telescopes, however they could be planned
and will not be part of a TOO program.  The SWIFT mission has been
selected for the Phase A Study and the final evaluation is due in
September 1999. Once selected the mission will fly in 2003.
\vskip -0.1cm

\section{Acknowledgments}
\vskip -0.4cm

I would like to thank the OAB staff who largely contributed to the
continuous development of the X-ray optics at OAB: Conconi, Ghigo,
Mazzoleni, Bergamini, Cantu`, Casiraghi, Crimi, Garignani, Salini and
Valtolina.  The Institutes involved in the hardware study of phase A
for {\bf WAXS/WFXT} were: Brera Astronomical Observatory (OAB),
Smithsonian Astrophysical Observatory (SAO), University of Leicester
(UL), Max-Planck-Institute für Extraterrestrische Physik (MPE). The
Italian Industries which partecipated: ALENIA, MEDIA-LARIO,
TELESPAZIO, OFFICINE GALILEO.  P.I: G. Chincarini (OAB);
co-PI:S. Murray (SAO); Co-I J. Trümper (MPE), A.  Wells (UL),
Telescope PI: O. Citterio (OAB); PM: G. Tagliaferri (OAB); PS:
S. Sciortino (OAPA).  Six panels headed by S. Sciortino coordinated
all the work for the science proposal. The panels were chaired by:
Colafrancesco (LSS), Chincarini (Clusters of galaxies), Zamorani
(AGN), Forman (Galaxies) Pallavicini (Stars) and Watson (Compact
Objects).  Contribution to this work came from Antonuccio-Delogu,
Arnaboldi, Bandiera, Bardelli, B\"ohringer, Bonometto, Borgani,
Campana, Catalano, Cavaliere, Conconi, Covino, Della Ceca, De Grandi,
De Martino, Fiore, Garilli, Ghisellini, Giacconi, Giommi, Girardi,
Giuricin, Governato, Guillout, Guzzo (who revised the final version of
the science proposal), Iovino, Israel, Lazzati, Le Fevre, Longo,
Maccacaro, Maccagni, Mardirossian, Matarrese, Micela, Molendi,
Molinari, Moscardini, Murray, Norman, Perola, Osborne, Pye, Ramella,
Randich, Robba, Rosati, Scaramella, Stella, Stewart, Tagliaferri,
Tozzi, Trinchieri, Vikhlinin, Vittorio, Wolter.  Members of the
following Institution expressed direct interest in the mission: ASI
(SDC), Bologna (OA), Catania (OA), Firenze (OA), Milano (IFCTR,UNI),
Napoli (OA), Padova (UNI,OA), Palermo (OA,UNI), Perugia (INFN), Roma
(OA-UNI), Trieste (OA), Marsiglia (CNRS,LAS), Baltimore (IHU),
Cambridge (MIT), Princeton (UNI), Munich (ESO,MPE), Copenhagen (DSRI).

Our involvement in the {\bf SWIFT} mission came following the proposal made
by Nick White to use the spare optics of the JETX Telescope (SRG
Mission) for a MIDEX proposal. The mission benefits of the extensive
experience of scientists at GSFC, Penn State University, Los Alamos
National Laboratory, Berkeley, Mullard Space Science laboratory (UK),
Leicester University, Brera Observatory, Italian Space Agency (ASI for
the Malindi Station). The SWIFT industrial partner Spectrum
Astro. Neil Gehrels (Goddard) is the Principal Investigator, Tim
Gehringer (Goddard) the Project Manager.  At present partecipate to
the mission the following co-Is/Lead representatives: N. White,
J. Nousek, L. Angelini, S. Barthelemy, T.  Cline, R. Corbet,
K. Jahoda, F. Marshall, R. Mushotxky, J. Norris, D.  Palmer,
A. Parsons, A. Smale, J. Tueller, W. Zhang, L. Whitlock, D.  Burrows,
G. Garmire, S. Horner, P. Mezaros, E. Feigelson, A. Wells, M.  Turner,
M. Ward, R. Willingale, K. Mason, M. Cropper, O. Citterio, G.
Chincarini, G. Tagliaferri, P. Caraveo, L. Stella, M. Vietri, E.
Fenimore, K. Hurley, F. Lebrun, J. Paul, B. Paczynski, F. Cordova, T.
Sasseen, L. Cominsky; and the following Associate Scientists
(Follow-up Team): M. Boer, M. Busby, A. Cimatti, M. della Valle,
A. Filippenko, D.  Frail, G. Ghisellini, P. Madau, B. Margon,
M. Metzger, H. pederson, P.  saracco, B. Schaefer, D. Schneider,
I. Smith, C. Stubbs, C. Thompson, F.  Vrba.

We refer to the proposal and web information for the various roles of
the participants to this fantastic mission. This is designed to serve
all of the international community, the data will be public and
working groups will be created for the follow-up research, interested
in this new and fascinating field of endeavor.

\vskip -0.4cm
\noindent
{\it http://www.merate.mi.astro.it/wfxt.html}

\noindent
{\it http://swift.gsfc.nasa.gov/}

\clearpage
\addcontentsline{toc}{section}{Index}
\flushbottom
\printindex

\end{document}